\documentclass[amt, manuscript]{copernicus}

\begin{document}

\nolinenumbers

\title{Numerical simulations and Arctic observations of surface wind effects on Multi-Angle Snowflake Camera measurements}


\Author[1,2]{Kyle E.}{Fitch}
\Author[3]{Chaoxun}{Hang}
\Author[1]{Ahmad}{Talaei}
\Author[1]{Timothy J.}{Garrett}

\affil[1]{Department of Atmospheric Sciences, University of Utah, Salt Lake City, 84112, USA}
\affil[2]{Department of Engineering Physics, Air Force Institute of Technology, Wright-Patterson Air Force Base, Ohio, 45433, USA}
\affil[3]{Department of Civil Engineering, Monash University, Clayton, 3168, Australia}




\correspondence{Kyle Fitch (kyle.fitch@us.af.mil)}

\runningtitle{Surface wind effects on MASC measurements}

\runningauthor{Fitch et al.}

\received{}
\pubdiscuss{} 
\revised{}
\accepted{}
\published{}


\firstpage{1}

\maketitle

\begin{abstract}

Ground-based measurements of frozen precipitation are heavily influenced by interactions of surface winds with gauge-shield geometry. The Multi-Angle Snowflake Camera (MASC), which photographs hydrometeors in free-fall from three different angles while simultaneously measuring their fall speed, has been used in the field at multiple mid-latitude and polar locations both with and without wind shielding. Here we show results of computational fluid dynamics (CFD) simulations of the airflow and corresponding particle trajectories around the unshielded MASC and compare these results to Arctic field observations with and without a Belfort double Alter shield. Simulations in the absence of a wind shield show a separation of flow at the upstream side of the instrument, with an upward velocity component just above the aperture, which decreases the mean particle fall speed by 55\%(74\%) for a wind speed of $5\,\unit{m\,s^{-1}}$($10\,\unit{m\,s^{-1}}$). MASC-measured fall speeds compare well with Ka-band Atmospheric Radiation Measurement (ARM) Zenith Radar (KAZR) mean Doppler velocities only when winds are light ($\leq5\,\unit{m\,s^{-1}}$) and the MASC is shielded. MASC-measured fall speeds that do not match KAZR measured velocities tend to fall below a threshold value that increases approximately linearly with wind speed but is generally $<0.5\,\rm{m\,s^{-1}}$. For those events with wind speeds  $\leq1.5\,\unit{m\,s^{-1}}$, hydrometeors fall with an orientation angle mode of $12\,\unit{^{\circ}}$ from the horizontal plane, and large, low-density aggregates are as much as five times more likely to be observed. We conclude that accurate MASC observations of the microphysical, orientation, and fall speed characteristics of snow particles require shielding by a double wind fence and restriction of analysis to events where winds are light ($\leq5\,\unit{m\,s^{-1}}$). Hydrometeors do not generally fall in still air, so adjustments to these properties' distributions within natural turbulence remain to be determined.
\end{abstract}


\introduction  
\label{intro}

Accurate measurement of snowfall is of importance to a wide range of scientific and public interests, including weather and climate prediction and monitoring \citep{Yang_etal_2005,rasmussen2012well,Theriault_etal_2015,Mekis_etal_2018}, hydrological cycles \citep{Yang_etal_2005,rasmussen2012well,Theriault_etal_2012,Mekis_etal_2018}, ecosystem research \citep{rasmussen2012well}, snowpack monitoring and disaster management \citep{Theriault_etal_2015,Mekis_etal_2018}, transportation \citep{Rasmussen_etal_2001,Theriault_etal_2012,Theriault_etal_2015}, agriculture \citep{Mekis_etal_2018}, and resource management \citep{Theriault_etal_2015,Mekis_etal_2018}. 

A persistent limitation of these studies is that catch-style precipitation gauges are prone to large uncertainties, especially when measuring snowfall in high winds -- a bias referred to as ``under-catch" \citep{Groisman_etal_1991,Groisman_Legates_1994,Goodison_etal_1998,Rasmussen_etal_2001,Yang_etal_2005}. A common remedy is to apply a correction based primarily on wind speed \citep{yang1993true,Rasmussen_etal_2001,rasmussen2012well,wolff2015derivation}, although hydrometeor type \citep{Theriault_etal_2012} and a dynamic drag coefficient \citep{colli2015improved} may also be considered. The correction is calculated by measuring the collection efficiency for a particular gauge or gauge-shield geometry, where collection efficiency is defined as the ratio of the gauge-measured precipitation rate to the best-estimate rate \citep{Theriault_etal_2012}. The Double Fence Intercomparison Reference (DFIR) is the standard reference, as determined by the World Meteorological Organization (WMO; \citeauthor{Goodison_etal_1998}, \citeyear{Goodison_etal_1998}); however, the DFIR has its own uncertainties which can lead to underestimation \citep{yang1993true} or even overestimation \citep{Theriault_etal_2015} of snowfall rates.  

Surface-based measurements of solid precipitation fall speed \citep{Garrett2014}, fall orientation \citep{Garrett2015,jiang2019shapes}, and size distributions \citep{Theriault_etal_2012} are all very sensitive to wind speed, with fall speed and size distribution having a strong influence on precipitation gauge collection efficiency \citep{Theriault_etal_2012,Theriault_etal_2015}. Accurate measurement of solid precipitation characteristics is important for constraining the densities and size distributions used in bulk microphysical parameterizations \citep[e.g.,][]{thompson2008explicit,morrison2015parameterization}. These parameters strongly influence bulk fall speed, highlighted by the Intergovernmental Panel on Climate Change (IPCC) as a critical factor for determining climate sensitivity \citep{IPCC_5AR_Ch9}. Likewise, knowledge of preferential hydrometeor orientation angles leads to the improved inference of hydrometeor shapes from backscattered polarimetric radar intensities \citep{vivekanandan1991rigorous,vivekanandan1994polarimetric,matrosov2005inferring,matrosov2015evaluations}, and these shapes combine with density to determine hydrometeor fall speeds \citep{bohm1989general}.

Past studies have typically combined airflow modeling and field observations to understand better the measurement error induced by winds and gauge geometry. Computational fluid dynamics (CFD) calculations characterize the wind velocity field and its interaction with various stationary objects in turbulent flows \citep{Moat2006,dehbi2008cfd,Ferrari2017}. \citet{Theriault_etal_2012} combined field observations and CFD simulations to better understand the scatter in collection efficiency as a function of wind speed for a Geonor, Inc. precipitation gauge located in a single Alter shield. Findings suggested that in addition to wind speed, the hydrometeor collection efficiency is a function of both hydrometeor type and size distribution. For example, hydrometeors such as graupel, with a relatively large density-to-surface-area ratio, fall faster and are collected more efficiently than large, low-density, aggregate-type hydrometeors. Additionally, \citet{colli2016collection1,colli2016collection2} compared shielded and unshielded gauge configurations using both time-averaged and time-dependent CFD simulations and found that a single Alter shield was effective in reducing the magnitude of turbulent flow above the gauge aperture. However, upwind shield deflector fins still produced turbulence that propagated into the collection area and generally reduced the collection efficiency.

One instrument that has received increased attention, but whose sampling characteristics have yet to be characterized in detail, is the Multi-Angle Snowflake Camera \citep[MASC;][]{Garrett2012}. The MASC system has overall dimensions of 43.5 cm x 58 cm x 21.5 cm \citep{stuefer2016multi} and observes particles falling into a ring-shaped collection area. The ring houses three cameras focused on a point at the ring center 10 cm away, with each camera separated by 36\unit{^\circ} \citep{Garrett2012}. A coupled system of directly opposing near-infrared emitters and detectors, vertically separated by 32 \unit{mm}, detect falling hydrometeors larger than approximately 0.1 \unit{mm} in maximum dimension. This triggers the cameras and three high-powered LEDs located directly above \citep{Garrett2014}. The time between triggers of the upper and lower emitter-detector pairs yields a fall speed. High-resolution images are captured at an exposure time of 1/25,000th of a second, sufficient to capture a vertical resolution of 40 \unit{\mu m} in a hydrometeor falling at 1 \unit{m\,s^{-1}} \citep{Garrett2012}.

The MASC system has helped to advance precipitation measurement by automating simultaneous high-resolution photography and fall speed measurement of falling hydrometeors from multiple angles, removing the need for tedious manual collection. Variables derived from the high-resolution images include those describing a hydrometeor's size, shape, fall orientation, and approximate riming degree \citep{Garrett2012,Garrett2014,Garrett2015}. As these hydrometeor properties are crucial for accurate numerical modeling and microwave scattering calculations, the MASC has been used at various polar and mid-latitude locations to constrain microphysical characteristics \citep{Garrett2012,Garrett2014,Garrett2015,grazioli2017measurements,kim2018observation,dunnavan2019shape,jiang2019shapes,kim2019quantitative,vignon2019microphysics}, improve radar-based estimates of snowfall rates \citep{gergely2016impact,cooper2017variational,schirle2019estimation}, automatically classify hydrometeors \citep{praz2017solid,besic2018unraveling,hicks2019method,leinonen2020unsupervised,schaer2020identification}, reconstruct particle shapes \citep{notaros2016accurate,kleinkort2017visual} and size distributions \citep{cooper2017variational,huang2017winter,schirle2019estimation}, and as ground truth comparisons for radar measurements \citep{bringi2017dual,gergely2017using,matrosov2017atmospheric,kennedy2018variations,oue2018toward,matrosov2019observational}.
Unlike more common precipitation gauges, the wind velocity field in the proximity of the MASC has not been simulated for various surface winds speeds, directions, or turbulence kinetic energies (TKE). 

Studies of hydrometeor behaviors using the MASC have shown, somewhat surprisingly, that frozen hydrometeor fall speeds are only weakly dependent on their size or shape, particularly under conditions of high turbulence intensity \citep{Garrett2014}. Prior studies had shown a much stronger dependence but had theoretically assumed or experimentally arranged for falling hydrometeors to settle in still air \citep{locatelli1974fall,bohm1989general}. MASC measurements led to a hypothesis that snow "swirls" in turbulent air in a manner that spreads particle fall speeds to both higher and lower values \citep{Garrett2014} -- an effect shown in prior work to be non-negligible in turbulent flows \citep{nielsen2007mean}. While the fact that snowflakes can just as readily move upwards as downwards is easily verified by any casual observations of a winter storm, it has remained unclear the extent to which the measurements of snowflake fall speed obtained by the MASC have been reflective of reality rather than some artifact of interactions of surrounding winds with the instrument body. 

In this study, we describe CFD simulations of hydrometeor-instrument interactions with specific application to the MASC. We compare these simulations to field observations of hydrometeor characteristics from a MASC located in the Arctic. The goal of this study is to better understand and characterize the influence of ambient wind speeds on MASC measurements of hydrometeor fall speed, fall orientation, and size distribution for both wind-shielded and unshielded configurations. 


\begin{figure}[H]
\includegraphics[width=1\textwidth]{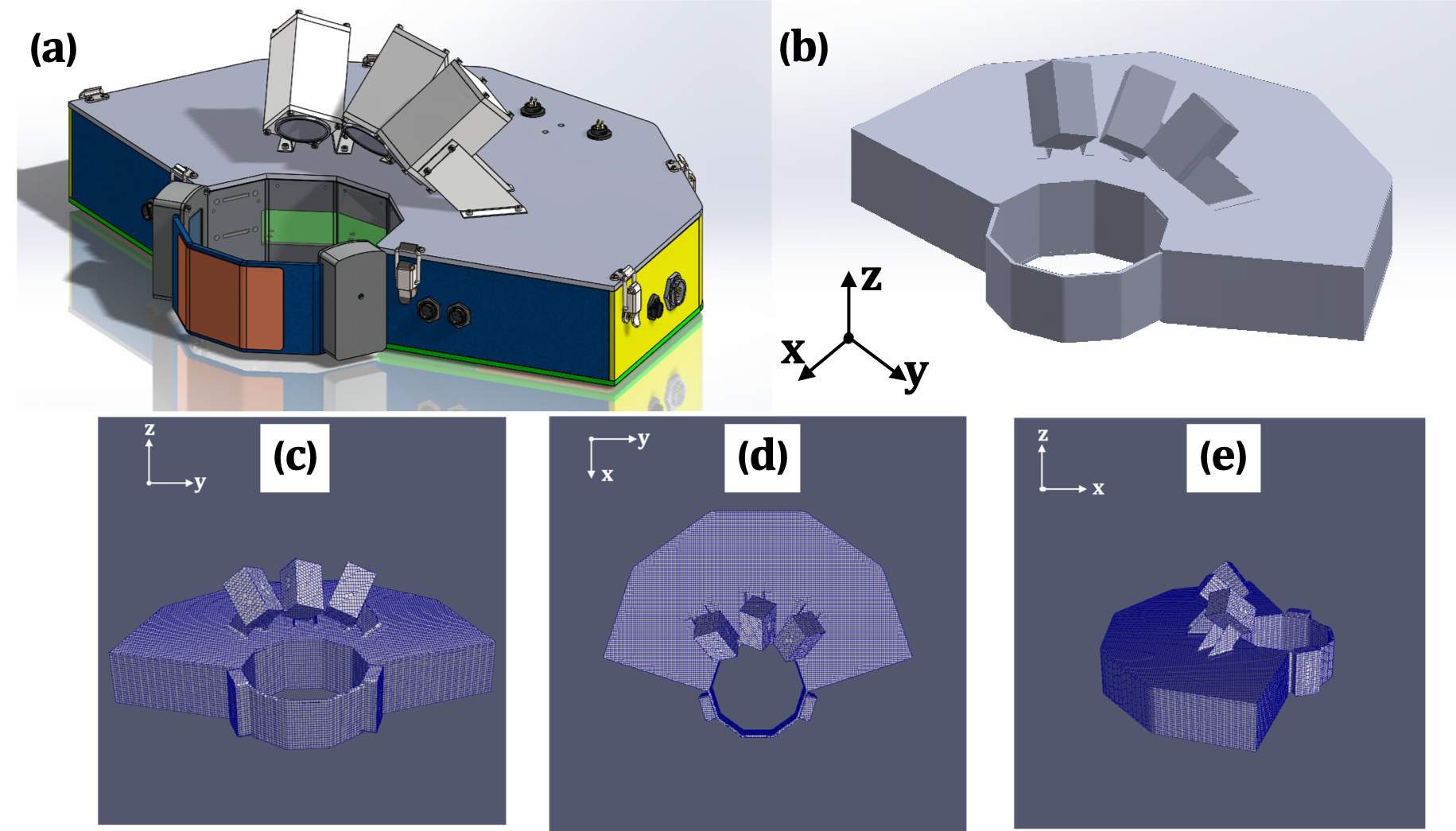}
\caption{(a) original MASC model as a Stereolithography (STL) file; (b) MASC model neglecting small-scale details (e.g., bolts, holes, patches, etc.); (c)--(e) snapped mesh on MASC in three viewing directions. }
\label{fig:mesh}
\end{figure}

\section{CFD simulations}\label{openFOAM}

To explore how ambient winds affect MASC measurements of fall speed, we use the OpenFOAM 4.1 tool \citep{jasak2007} for CFD calculations of falling particles and winds interacting with the MASC body. OpenFOAM is an open-source CFD toolbox based on C++ libraries and codes designed to solve complex flow dynamics problems \citep{jasak2007,Chen2014,Greenshields2015}. The solver uses the factorized finite volume method (FVM) with the Semi-Implicit Method for Pressure Linked Equations (SIMPLE) algorithm \citep{caretto1973two} to solve the Navier--Stokes equations. The $k$--$\omega$ Shear Stress Transport (SST) model is utilized in this study to solve the turbulence closure problem due to its capability to capture the flow separation near objects through the viscous sub-layer, without additional wall functions \citep{menter1993zonal}. We combine the incompressible, robust \textit{simpleFOAM} solver for steady incompressible turbulent flows \citep{balogh2012rans, higuera2014three} with the \textit{solidParticle} and \textit{solidParticleCloud} classes to study the motions of  particles \citep{iudicianilagrangian}.  

To study particle-air interactions, the first step is to determine the two-phase flow type. The ratio between the average inter-particle distance and the particle diameter is estimated. Provided the ratio is $\gtrsim 100$, the flow can be treated as a dilute dispersed system, and one-way coupling -- wherein the particles do not collide with each other and also do not affect the flow field -- can be assumed \citep{elghobashi1994predicting}. The OpenFOAM \textit{blockMesh} and \textit{snappyHexMesh} tools are applied here to generate a mesh around the complex physical geometry of the MASC instrument \citep{gisen2014generation}. The \textit{snappyHexMesh} utility automatically generates 3D meshes containing hexahedra and split-hexahedra efficiently. Figure \ref{fig:mesh}(c--e) shows the MASC mesh for different viewing angles. Spatial and temporal parameters are provided in Table \ref{table:parameter}.

\begin{table}[t]
\caption{Domain size and fluid and particle properties of simulations }
\begin{tabular}{l c r}
\tophline
\middlehline
\textbf{Domain Dimensions}           &               & \\
Width (x--dir)               & 4 \unit{m}    & \\
Transverse thickness (y--dir)& 4 \unit{m}    & \\
Height (z--dir)              & 10 \unit{m}   & \\
Grid (x $\, \times \,$ y $\, \times \,$ z)   & 16 $\times$ 16 $\times$ 40 & \\
\textbf{Particle properties}         &               & \\
Number of particles         & 400           & \\
Diameter ($D_p$)            & 2 \unit{mm}   & \\
Density ($\rho_p$)          & 50 \unit{kg\, m^{-3}} & \\
\textbf{Fluid properties (at 0 $^{\circ}$C) }          &               & \\
Viscosity ($\mu$)           & 1.34 $\times 10^{-5}$ \unit{m^2\,s^{-1}}& \\
Density ($\rho_f$)          & 1.284 \unit{kg\, m^{-3}}   & \\
\bottomhline
\label{table:parameter}
\end{tabular}
\belowtable{} 
\end{table}

For the simulation of hydrometeors in the atmosphere, we track spherical particles of mass $m_p$, diameter $D_p$, and area $A_p$ within a Lagrangian framework, where the Eulerian fluid velocity field $\vec{v}_f=v_{f_x}\hat{x}+v_{f_y}\hat{y}+v_{f_z}\hat{z}$ is interpolated from nearby grid points at the position of the particle to compute the instantaneous particle drag. The particle velocity $\vec{v}_p$ is calculated at each time step by assuming that the particle's Reynolds number $Re_p$ is greater than unity, which gives a reduced form of the Maxey--Riley equation of motion \citep{maxey1983equation}: 
\begin{equation} 
m_p\frac{d{\vec{v}}_p}{dt} = m_p{\vec{g}}- \frac{1}{2}C_D(Re) \rho_f A_p |\vec{v}_p(t)-\vec{v}_f(t)| \left( \vec{v}_p(t)-\vec{v}_f(t) \right) \label{eq:main_force_eq} 
\end{equation} 
where the drag coefficient $C_D(Re)$ is a function of the relative Reynolds number $Re=\frac{(v_p-v_f)D_p}{\mu}$, $\rho_f$ is the fluid density, and $g$ is the gravitational constant. Particles measured with the MASC had a median $Re_p$ of 108, with $95\%$ of the values in the range of $40<Re<360$.

In simulations of the response of the particles to horizontal winds in the vicinity of the MASC, the particles are evenly distributed on a 20$\times$20 grid with 1 \unit{mm} spacing in the x--direction and  2 \unit{mm} spacing in the y--direction. The particles fall downward at an initial velocity of $1\,\unit{m\,s^{-1}}$ from a height of $3\,\unit{m}$ above the MASC in the $-z$ direction under the force of gravity, reaching an average terminal velocity of $1.05\,\unit{m\,s^{-1}}$ well before encountering flows perturbed by the MASC.

\begin{figure}[t]
\includegraphics[width=1\textwidth]{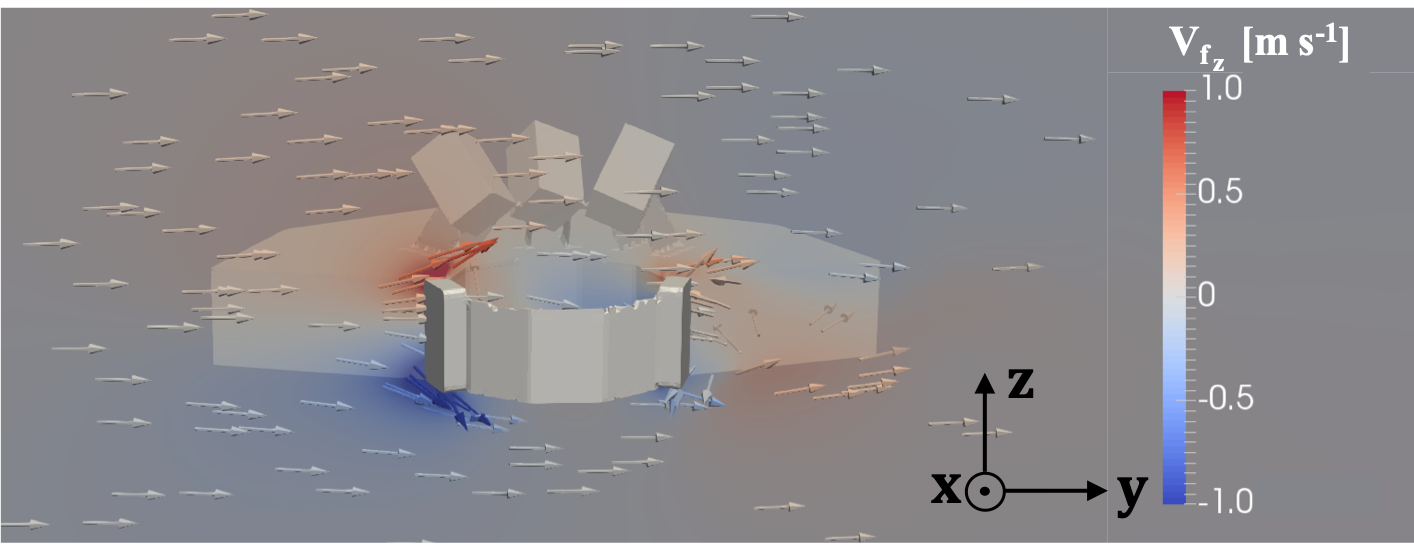}
\caption{Simulated wind field around the MASC with undisturbed winds set at $1\,\unit{m\,s^{-1}}$ towards the positive y--direction. Color represents the vertical wind speed $v_{f_z}$, and arrows show wind vectors on the y--z plane. The cross-section is in the middle of the aperture on the y--z plane, and x--positive points out of the page. }
\label{fig:windField}
\end{figure}

Figure \ref{fig:windField} shows interactions of a horizontal flow in the +y direction of 1 \unit{m\,s^{-1}} with the MASC body. There is a clear separation of flow on the upstream side of the aperture, a relatively large upward component above the aperture, and a smaller downward component within the aperture. 

\begin{figure}[t]
\includegraphics[width=1\textwidth]{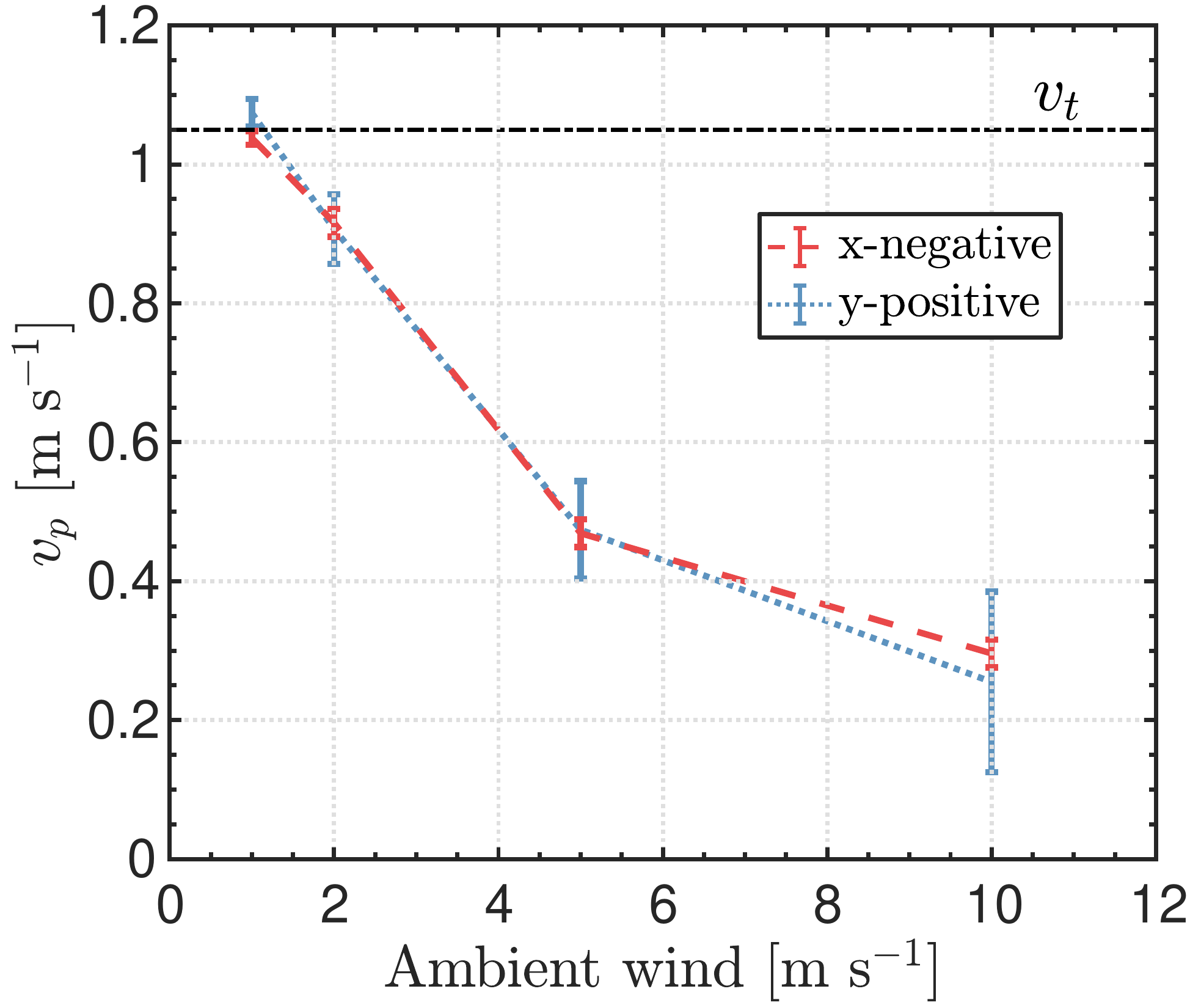}
\caption{Mean fall speed of particles $v_p$ as a function of ambient wind speed. Error bars represent the standard deviation of all the particles at each ambient wind speed. x--negative and y--positive represent the wind pointing towards $-x$ and $+y$ directions (see Fig. \ref{fig:mesh}(d)), respectively. Terminal fall speed $v_t$ is included for comparison, and the initial TKE is 1 \unit{m^2\,s^{-2}}.}
\label{fig:windSpd}
\end{figure}

The response of particles to these perturbations for horizontal winds in both the $-x$ and $+y$ directions is shown in Fig. \ref{fig:windSpd}. There is low sensitivity to wind direction, but the mean particle fall speed within the MASC aperture decreases from $1.07(1.04)\,\unit{m\,s^{-1}}$ to $0.30(0.26)\,\unit{m\,s^{-1}}$ as the ambient wind speed increases from 1 to $10\,\unit{m\,s^{-1}}$ (Table \ref{table:v_p}).

\begin{figure}[t]
\includegraphics[width=1\textwidth]{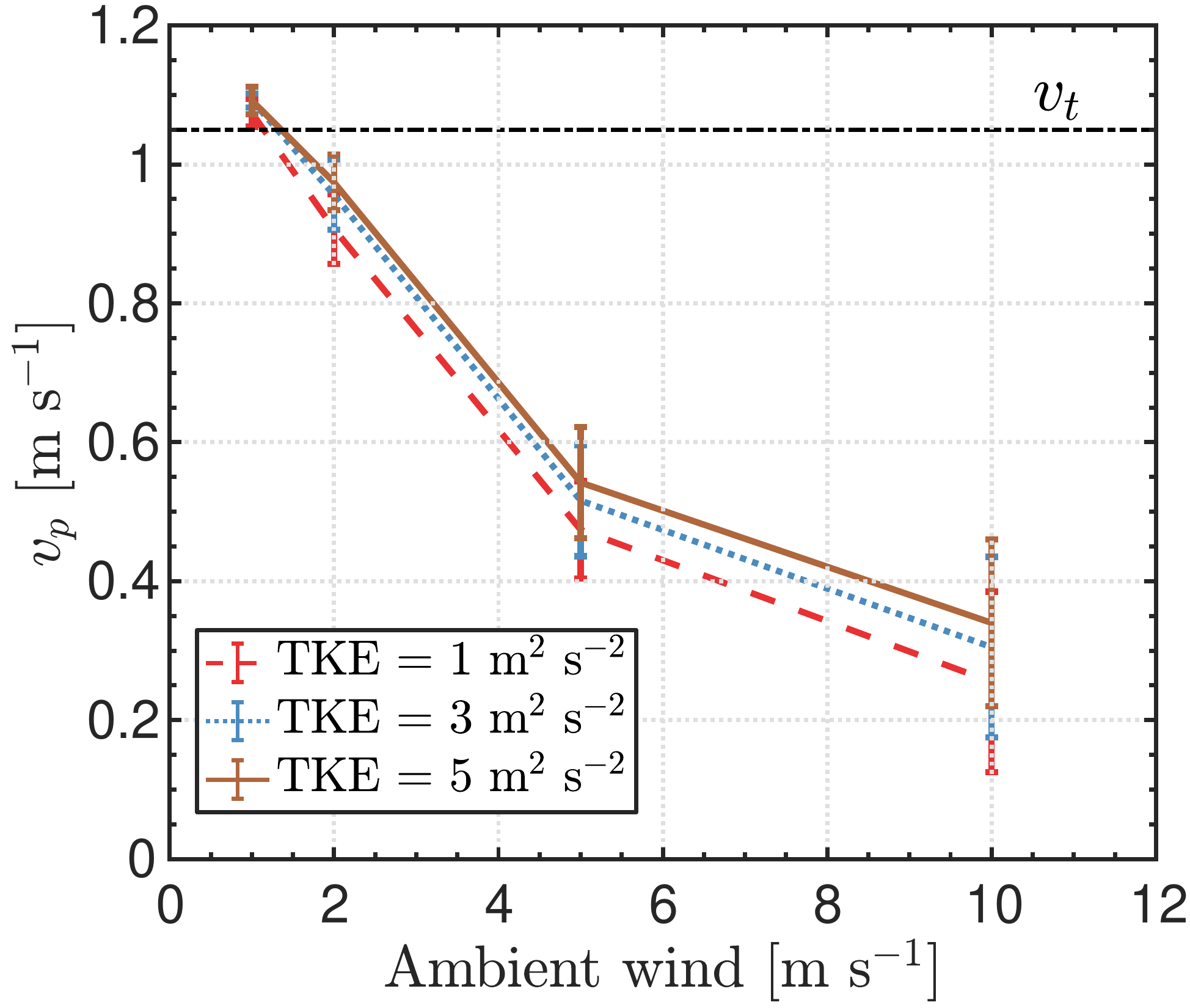}
\caption{Mean fall speed $v_p$ of particles versus ambient wind speed for different values of initial TKE. Terminal fall speed $v_t$ is included for comparison.}
\label{fig:tke}
\end{figure}

The influence of ambient turbulent intensity expressed as $TKE = \frac{1}{2} \big( \overline{{v'_{f_x}}^2} + \overline{{v'_{f_y}}^2} + \overline{{v'_{f_z}}^2}\big)$ was calculated for $TKE= 1$, 3, and 5 \unit{m^2\,s^{-2}}, where the perturbation velocity $v'_f$ is the difference between the instantaneous and average velocities of the atmospheric flow. These TKE values are used as initial conditions in the $k$--$\omega$ closure model, which determines the shear stress, which in turn is used in the momentum budget equation. Figure \ref{fig:tke} shows that for a wind speed of $10\,\unit{m\,s^{-1}}$, the mean particle fall speed is 24\% lower for an initial value of $TKE=1\,\unit{m^2\,s^{-2}}$ than it is for $TKE=5\,\unit{m^2\,s^{-2}}$ (Table \ref{table:v_p}).

\begin{table}[t]
\caption{Mean particle fall speed for various wind directions, wind speeds, and TKE values. The terminal fall speed is 1.05 \unit{m\,s^{-1}} in all runs.}
\begin{tabular}{ccccc}
\tophline
\textbf{Ambient\, wind} & $\mathbf{1\,\unit{m\,s^{-1}}}$ & $\mathbf{2\,\unit{m\,s^{-1}}}$ & $\mathbf{5\,\unit{m\,s^{-1}}}$ & $\mathbf{10\,\unit{m\,s^{-1}}}$ \\
\middlehline
\textbf{Wind Direction} & & & & \\
x-negative &  1.07\,\unit{m\,s^{-1}} &  0.92\,\unit{m\,s^{-1}} &  0.47\,\unit{m\,s^{-1}} &  0.30\,\unit{m\,s^{-1}} \\
y-positive &  1.04\,\unit{m\,s^{-1}} &  0.91\,\unit{m\,s^{-1}} &  0.47\,\unit{m\,s^{-1}} &  0.26\,\unit{m\,s^{-1}} \\
\middlehline
\textbf{TKE} & & & & \\
 1\,\unit{m^2\,s^{-2}} &  1.07\,\unit{m\,s^{-1}} &  0.91\,\unit{m\,s^{-1}} &  0.47\,\unit{m\,s^{-1}} &  0.26\,\unit{m\,s^{-1}} \\
 3\,\unit{m^2\,s^{-2}} &  1.09\,\unit{m\,s^{-1}} &  0.96\,\unit{m\,s^{-1}} &  0.52\,\unit{m\,s^{-1}} &  0.31\,\unit{m\,s^{-1}} \\
 5\,\unit{m^2\,s^{-2}} &  1.09\,\unit{m\,s^{-1}} &  0.98\,\unit{m\,s^{-1}} &  0.54\,\unit{m\,s^{-1}} &  0.34\,\unit{m\,s^{-1}} \\
\bottomhline
\end{tabular}
\belowtable{} 
\label{table:v_p}
\end{table}

\begin{figure}[t]
\includegraphics[width=1\textwidth]{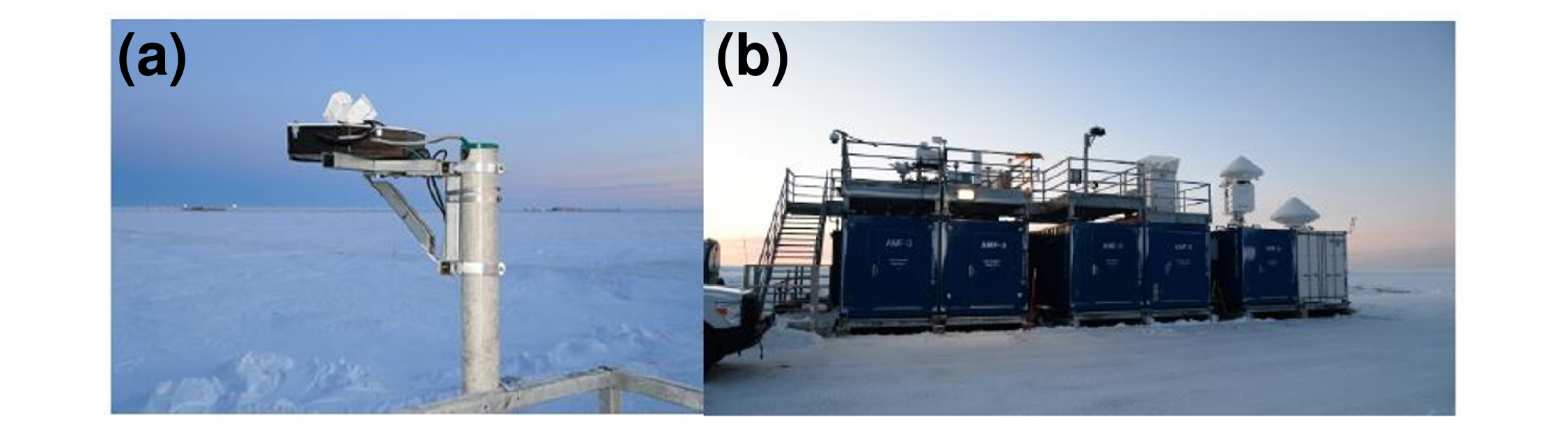}
\caption{(a) Unshielded MASC configuration at the Third ARM Mobile Facility (AMF3), Oliktok Point, Alaska. (b) Ground-level view of the MASC on top of a group of shipping containers. This was the MASC configuration from initial deployment in February 2015 through 21 August 2016. Image courtesy of the U.S. Department of Energy Atmospheric Radiation Measurement (ARM) user facility.}
\label{fig:noshield_masc}
\end{figure}


\section{Hydrometeor observations}\label{obs}

\subsection{Methods}
\label{obs_methods}
Processing of MASC imagery consists of distinguishing foreground pixels from background to define the region of interest (ROI) and then fitting the ROI with a bounding ellipse \citep{shkurko_etal_2018}. The ellipse's major axis is defined as the maximum dimension $D_{max}$ for each image. The absolute value of the angle between $D_{max}$ and the local horizontal plane is the orientation angle $\theta$ \citep{Garrett2012,Garrett2014,Garrett2015,shkurko_etal_2018}. A complexity parameter $\chi$ is used to distinguished riming classes \citep{Garrett2014}. Here we use $\chi\leq1.35$ to identify heavily rimed graupel, $1.35<\chi\leq2.00$ for moderate riming, and $\chi>2.00$ indicates sparsely-rimed aggregates. We note that a value of 1.75 was used to distinguish moderately rimed particles from aggregates for Utah snow measurements in \citet{Garrett2014}, with the observation that the value is subjectively determined by visual inspection of hydrometeor images and varies with location. Mean values of fall speed $v_p$, $D_{max}$, $\theta$, and $\chi$ from all three images are used for each particle.

\begin{figure}[t]
\includegraphics[width=1\textwidth]{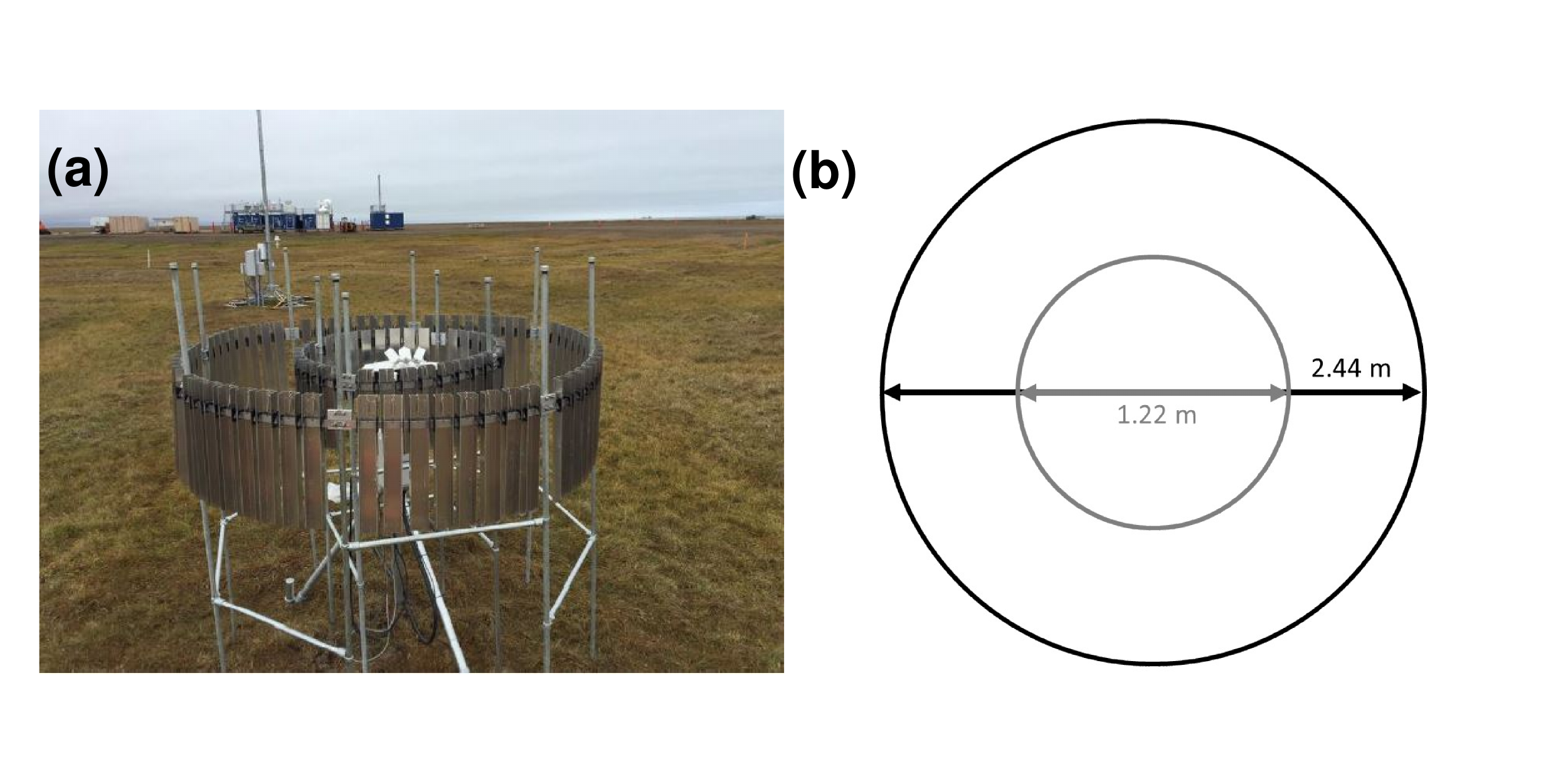}
\caption{(a) The MASC was relocated to ground level and placed inside a Belfort double Alter shield on 22 August 2016 (field site photograph courtesy of Martin Stuefer). (b) The shield consists of inner and outer fences with diameters of 1.22 \unit{m} and 2.44 \unit{m}, respectively.}
\label{fig:shield_masc}
\end{figure}
 
A MASC was installed at the Department of Energy's Third Atmospheric Radiation Measurement (ARM) Mobile Facility (AMF3), Oliktok Point, Alaska, in February 2015. The initial deployment was atop a group of shipping containers with no wind shield (Fig. \ref{fig:noshield_masc}). On 22 August 2016, the MASC was relocated to ground level and placed inside of a Belfort Model 36001 Double Alter Wind Shield (Fig. \ref{fig:shield_masc}). The central camera was pointed in the east-northeasterly direction \citep{jiang2019shapes}, with surface wind observations showing this to be the predominant wind direction for the present study. The inner(outer) fence of the shield is 1.22(2.44) $\unit{m}$ in diameter, with 32(64) deflector fins that are each 46(61) $\unit{cm}$ in length. Observations used here include both unshielded and shielded configurations, spanning a 33-month period from 29 November 2015 to 28 August 2018 \citep{MASC_data}. Raw data and images were processed with a local University of Utah processing suite called \textit{mascpy} \citep{HIVE_mascpy,HIVE_wecode}, similar to that described in \citet{shkurko_etal_2018}.

To complement MASC observations and characterize the influence of ambient wind speed on MASC measurements, surface wind measurements from a traditional meteorological ground suite \citep{MET_handbook,MET_data} were matched to MASC hydrometeors by calculating a mean wind speed for the 1 minute period leading up to the observation time corresponding to each hydrometeor. In addition to the quality control checks listed in \citet{shkurko_etal_2018}, a surface temperature threshold of $<2\,\unit{^{\circ}C}$ was used to exclude liquid hydrometeors, which are occasionally misidentified by the \textit{mascpy} algorithm.

For a ground truth hydrometeor fall speed, mean Doppler velocity was calculated from the volume of scattering hydrometeors detected by a co-located Ka-band ARM Zenith-pointing Radar (KAZR). At a vertical resolution of 30 m, the KAZR produces measurements of the first three moments of the Doppler spectrum: reflectivity, mean Doppler velocity, and spectrum width \citep{widener2012ka,oue2018toward}. The Doppler velocity signal has a resolution of 0.05 \unit{m\,s^{-1}} \citep{oue2018toward} and consists of both larger particle fall speeds and the vertical air motions traced by smaller particles \citep{shupe2008vertical}. Using only Doppler velocity measurements originating from below cloud base, we isolate the signal of the larger, precipitation-sized hydrometeors, while acknowledging the relatively small bias of Doppler broadening from turbulence and wind shear \citep{shupe2008vertical}. Both mean Doppler velocity and cloud base height were retrieved from the ARM's KAZR Active Remote Sensing of CLouds (ARSCL) Value-Added Product \citep{KOLLIAS_data,Clothiaux_etal_2000}.

Results are presented here in the form of probability density function (PDF) estimates, calculated by normalizing the frequency $N_i$ of each histogram bin $i$ of width $\Delta x$ (equally spaced bins), such that $PDF\simeq \frac{N_i}{N_t \Delta x}$. Here $N_t=\sum N_i$ is the total number of observations, and $x$ is the variable of interest. The resulting PDF estimates were adjusted using a Gaussian kernel smoothing function. For distributions of $D_{max}$, the exponential slope parameter $\lambda$ is computed using a linear least squares regression from the peak of the log-linear distribution through the tail. 


\begin{figure}[H]
\centering
\includegraphics[width=1\textwidth]{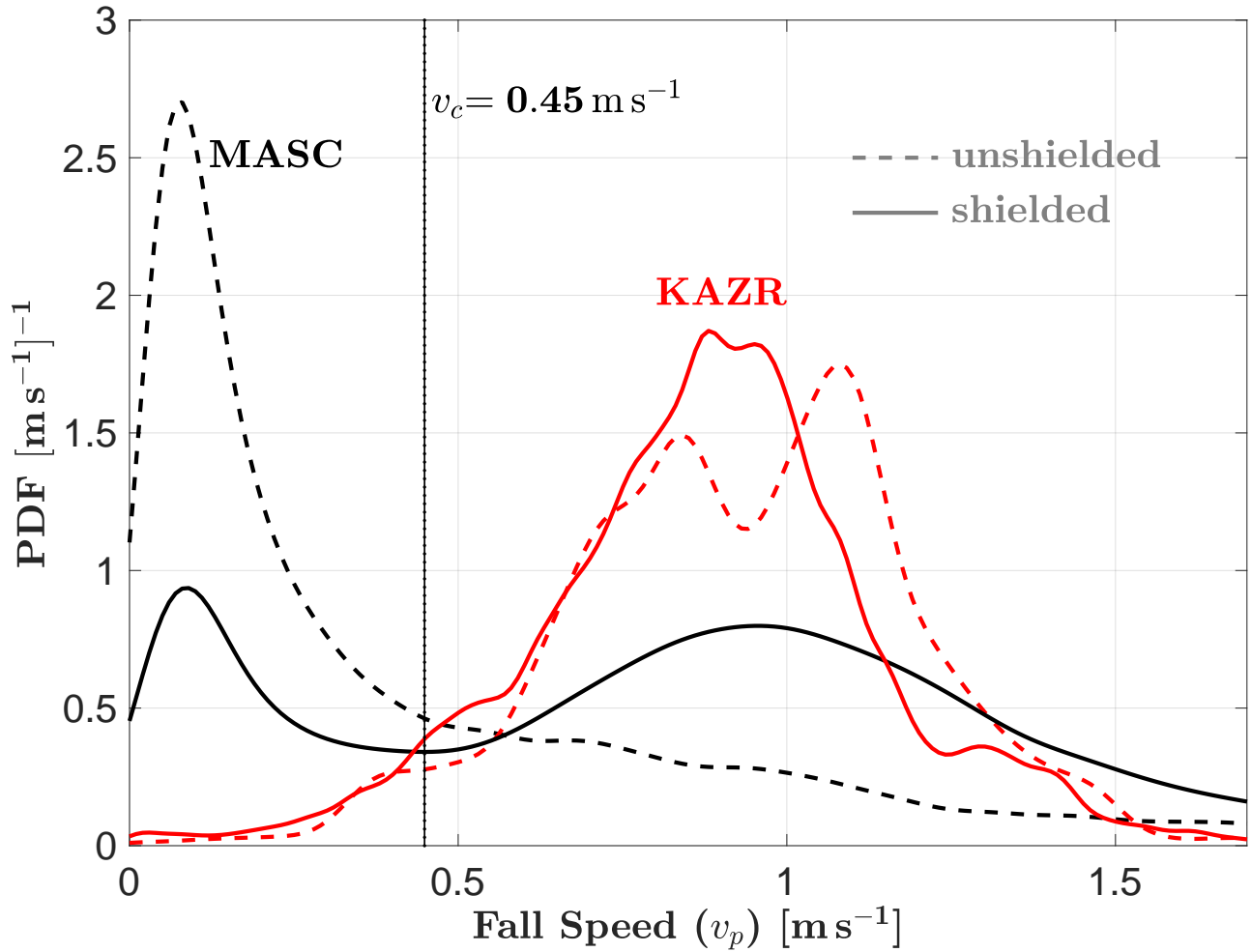}
\caption{Comparison of fall speed $v_p$ probability density function (PDF) estimates from MASC and KAZR measurements, both with and without wind shielding of the MASC. KAZR fall speeds are determined from the mean Doppler velocity below cloud base (positive downward, see Sect. \ref{obs_methods} for details). The cutoff fall speed $v_c$ marks the location of the local minimum separating the two modes of the shielded MASC distribution.}
\label{fig:v_pdf}
\end{figure}

\subsection{Observations of fall speed}
\label{obs_fall_speed}
Distributions of MASC-measured particle fall speed $v_p$, both with and without a wind shield, are compared to coincident measurements from the KAZR in Fig. \ref{fig:v_pdf}. The KAZR-measured fall speed mode is $\sim1\,\unit{m\,s^{-1}}$, while the MASC-measured fall speed distribution has a mode of 0.08 \unit{m\,s^{-1}} for both the shielded and unshielded cases. However, the shielded MASC fall speed distribution has a second mode at 0.96 \unit{m\,s^{-1}}, similar to the location of the KAZR mode. Notably, a low-speed mode was not observed in the KAZR measurements despite its velocity resolution of 0.05 \unit{m\,s^{-1}}.

\begin{figure}[H]
\centering
\includegraphics[width=0.80\textwidth]{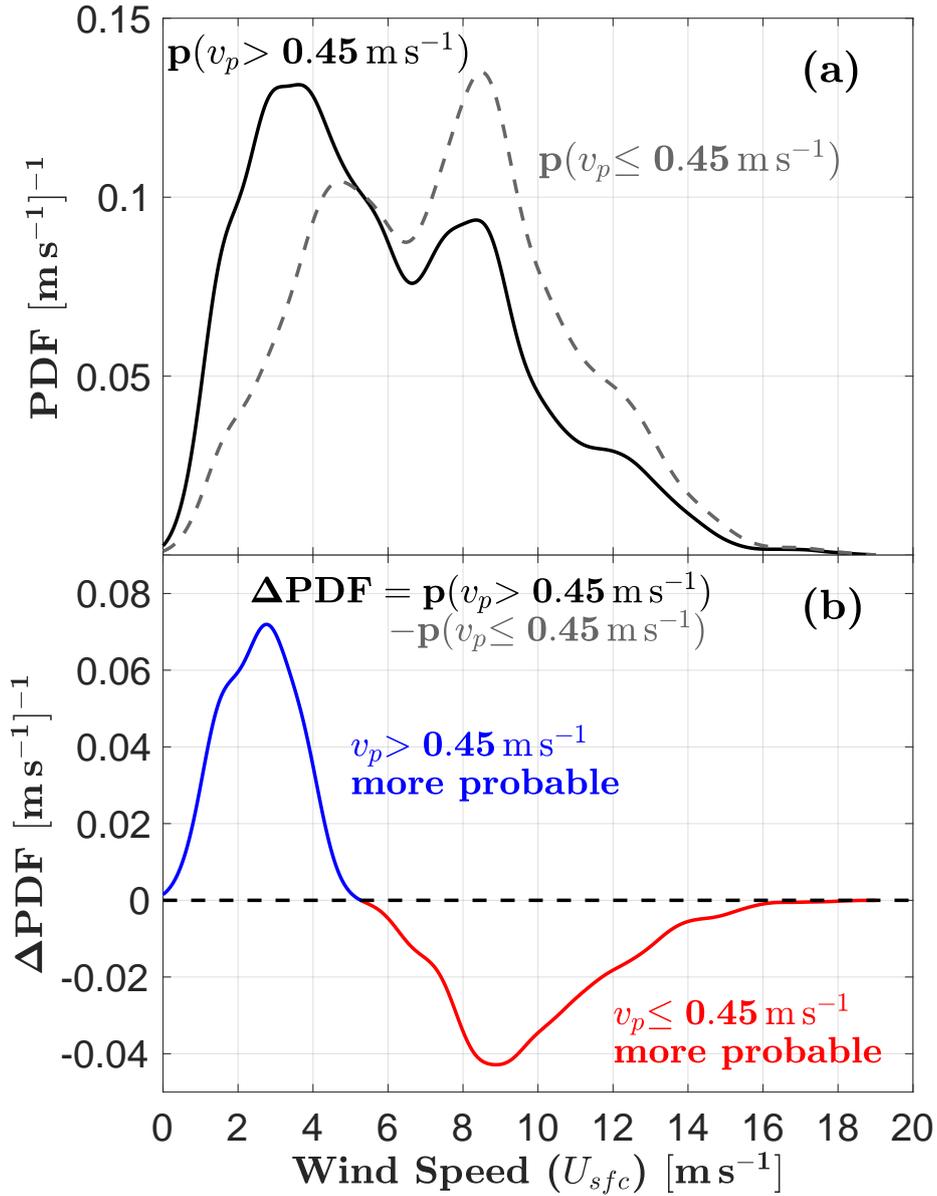}
\caption{(a) Comparison of, and (b) difference between, estimates of surface wind speed $U_{sfc}$ probability density functions (PDFs) for the high ($v_p>0.45\,\unit{m\,s^{-1}}$) and low ($v_p\leq0.45\,\unit{m\,s^{-1}}$) fall speed modes of the shielded MASC fall speed distribution from Fig. \ref{fig:v_pdf}. $\Delta PDF>0$ means the probability of $v_p>0.45\,\unit{m\,s^{-1}}$ is greater.}
\label{fig:delta_pdf}
\end{figure}

The shielded MASC fall speed distribution deviates substantially from the corresponding KAZR distribution for fall speeds below $0.45\,\unit{m\,s^{-1}}$. This is the location of the local minimum separating the two modes of the shielded MASC fall speed distribution and is defined from here on as the cutoff fall speed $v_c$: the fall speed below which MASC measurements are assumed to be erroneous. The fall speed distribution can therefore be divided into two parts: $v_p>v_c$ and $v_p\leq v_c$. 

\begin{figure}[H]
\centering
\includegraphics[width=1\textwidth]{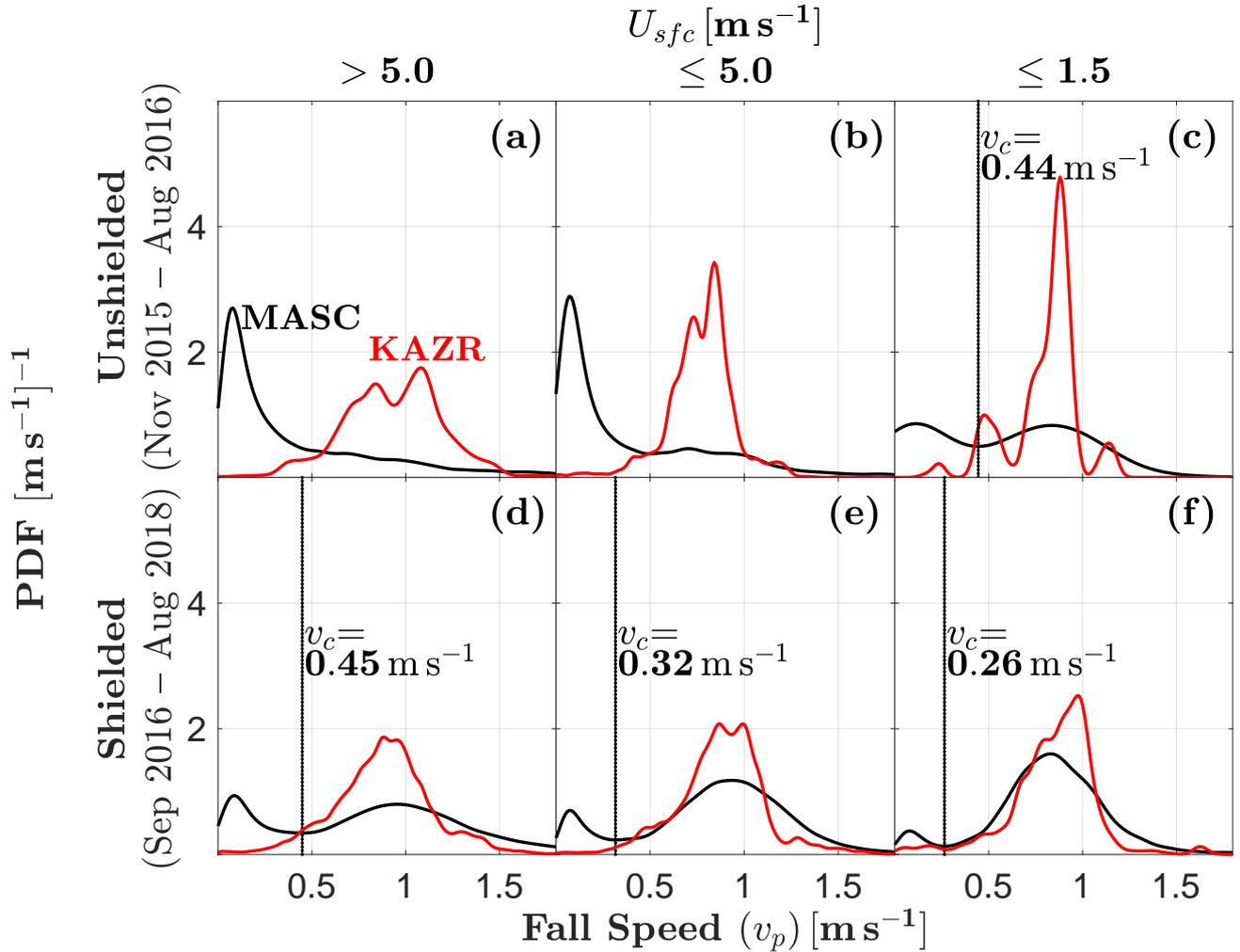}
\caption{Comparison of MASC hydrometeor fall speed and KAZR mean Doppler fall speed distributions for (a)--(c) unshielded and (d)--(f) shielded MASC measurements. Surface wind speeds $U_{sfc}$ decrease from left to right. Where MASC PDFs are bimodal, the vertical line marks the cutoff fall speed $v_c$, indicating the location of the local minimum of the PDF separating the two modes. The number of observations for each case is listed in Table \ref{table:counts}. The terms ``shielded" and ``unshielded" refer only to the MASC. }
\label{fig:v_vd_comp}
\end{figure}

To examine the influence of surface wind speeds on MASC fall speed measurements, Fig. \ref{fig:delta_pdf} shows PDF estimates of wind speed $U_{sfc}=\sqrt{v_{f_x}^2+v_{f_y}^2}$ for the two separate parts of the shielded MASC fall speed distribution from Fig. \ref{fig:v_pdf}. From the difference (Fig. \ref{fig:delta_pdf}(b)), it is apparent that the high-speed mode of $v_p>0.45\,\unit{m\,s^{-1}}$ is more likely to be observed when $U_{sfc}<5\,\unit{m\,s^{-1}}$. This matches well with the simulated fall speeds in wind speeds of $\leq5\,\unit{m\,s^{-1}}$ from Table \ref{table:v_p} and Figs. \ref{fig:windSpd} and \ref{fig:tke}, although the simulation did not include a wind fence.

\begin{figure}[t]
\centering
\includegraphics[width=1\textwidth]{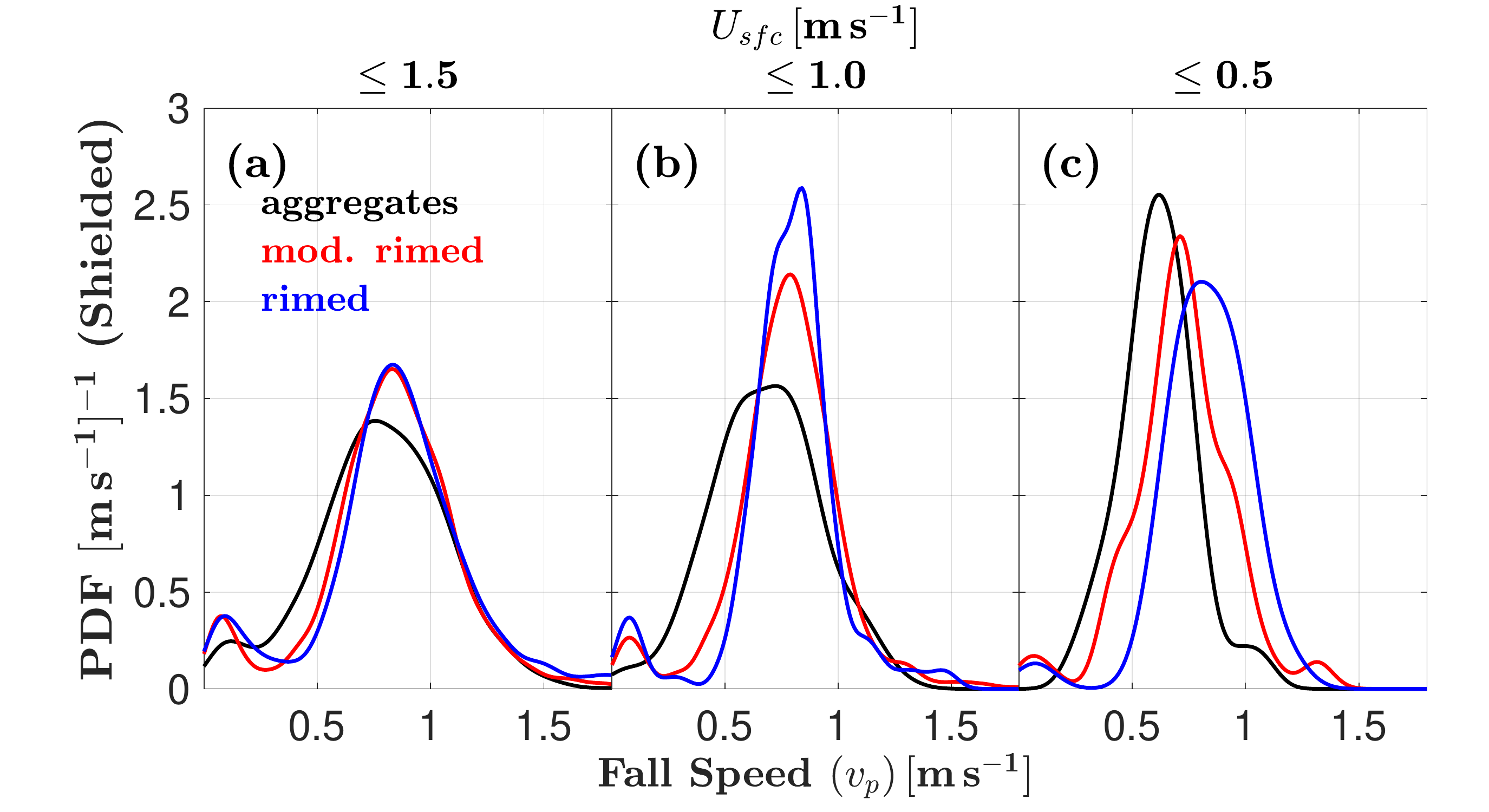}
\caption{Probability density function (PDF) estimates for shielded MASC fall speed $v_p$ for very light wind speeds and hydrometeors divided into three riming classes: sparsely-rimed aggregates, moderately rimed, and rimed. }
\label{fig:v_chi}
\end{figure}

Figure \ref{fig:v_vd_comp} compares MASC and KAZR fall speed distributions as a function of $U_{sfc}$, again both with and without wind shielding of the MASC. Qualitatively, the agreement between the MASC and KAZR distributions is maximized for shielded MASC measurements with light winds ($U_{sfc}\leq1.5\,\unit{m\,s^{-1}}$), where only 7\% of measured fall speeds are lower than the $v_c$ threshold of $0.26\,\unit{m\,s^{-1}}$ (Fig. \ref{fig:v_vd_comp}(f)). When separated by riming class, shielded MASC fall speed distributions show discernible differences only for the lightest winds. This is most apparent for $U_{sfc}\leq0.5\,\unit{m\,s^{-1}}$ (Fig. \ref{fig:v_chi}(c)), where the most heavily rimed particles ($\chi\leq1.35$) tend to exhibit the highest fall speeds. Particle counts corresponding to Figs. \ref{fig:v_vd_comp} and \ref{fig:v_chi} are listed in Table \ref{table:counts}.

\begin{table}[t]
\caption{Number and percentage of observed hydrometeors in each wind shielding case, surface wind speed $U_{sfc}$ category, and riming class. Percentages may not add to precisely 100\% due to rounding.}
\begin{tabular}{cccccc}
\tophline
 &  &  & $U_{sfc}$ &  &  \\
\textbf{Category} & $\mathbf{>5\,\unit{m\,s^{-1}}}$ & $\mathbf{\leq5\,\unit{m\,s^{-1}}}$ & $\mathbf{\leq1.5\,\unit{m\,s^{-1}}}$ & 
$\mathbf{\leq1.0\,\unit{m\,s^{-1}}}$ & $\mathbf{\leq0.5\,\unit{m\,s^{-1}}}$ \\
\middlehline
\textbf{No Wind Shield} & \textbf{2,249} & \textbf{5,097} & \textbf{460} 
& \textbf{167} & \textbf{32}\\
Aggregates & 176 (8\%) & 1,522 (30\%) & 67 (15\%) & 15 (9\%) & 5 (16\%) \\
Moderately Rimed & 1,209 (54\%) & 2,891 (57\%) & 315 (68\%) & 115 (69\%) & 14 (44\%) \\
Rimed & 864 (38\%) & 684 (13\%) & 78 (17\%) & 37 (22\%) & 13 (41\%) \\
\middlehline
\textbf{Wind Shield} & \textbf{85,151} & \textbf{58,939} & \textbf{5,730} 
& \textbf{1,372} & \textbf{161} \\
Aggregates & 15,320 (18\%) & 11,304 (19\%) & 1,299 (23\%) & 302 (22\%) & 41 (25\%) \\
Moderately Rimed & 47,147 (55\%) & 35,820 (61\%) & 3,477 (61\%) & 855 (62\%) & 86 (53\%) \\
Rimed & 22,684 (27\%) & 11,815 (20\%) & 954 (17\%) & 215 (16\%) & 34 (21\%) \\
\bottomhline
\end{tabular}
\belowtable{} 
\label{table:counts}
\end{table}


\subsection{Observations of orientation, maximum dimension, and riming degree}

\begin{figure}[t]
\centering
\includegraphics[width=1\textwidth]{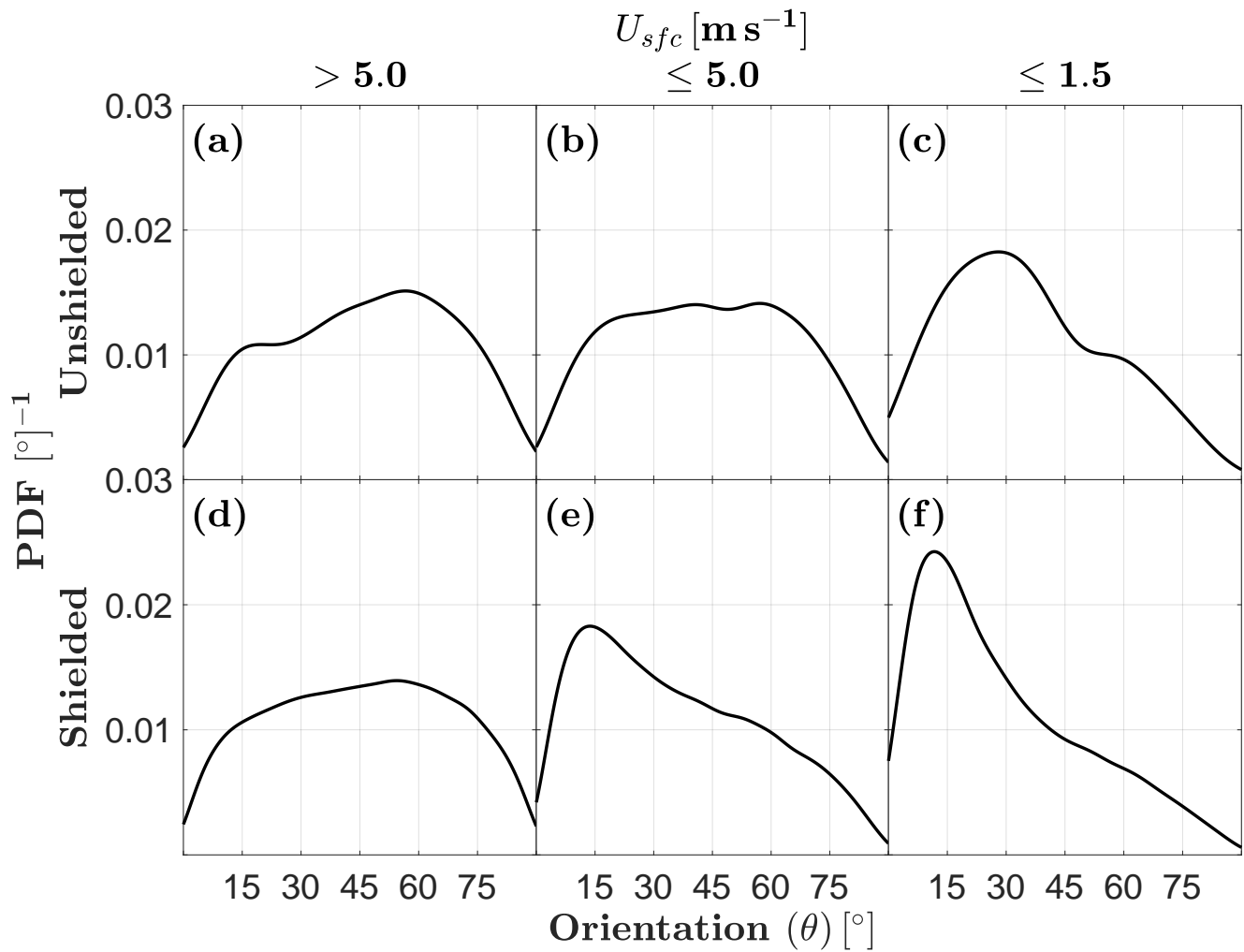}
\caption{Probability distribution function (PDF) estimates of MASC-observed orientation angle $\theta$ as a function of surface wind speed $U_{sfc}$ for both shielded and unshielded configurations.}
\label{fig:theta}
\end{figure}

\begin{figure}[t]
\centering
\includegraphics[width=1\textwidth]{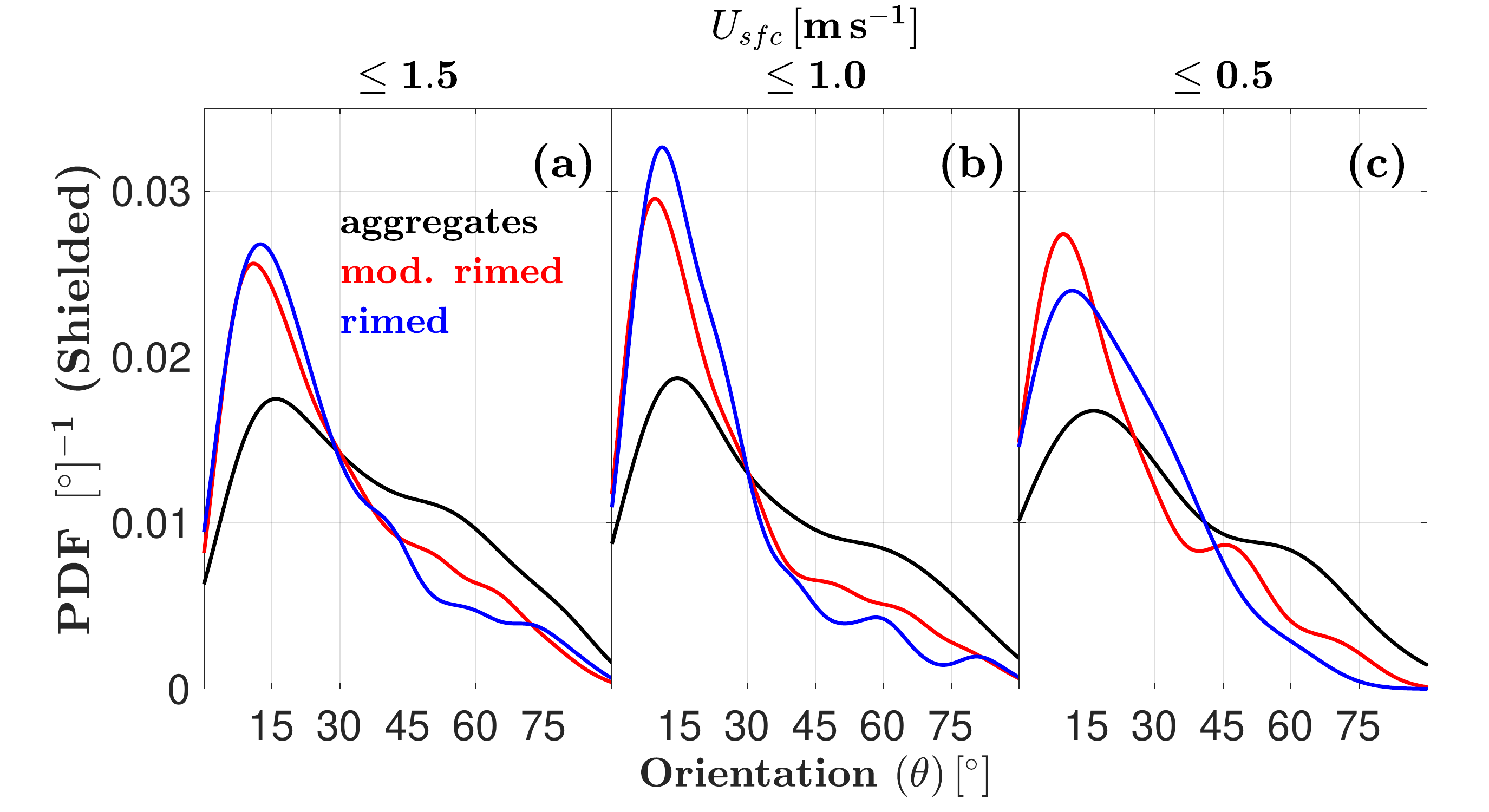}
\caption{As in Fig. \ref{fig:theta} but with lighter winds and hydrometeors divided into three riming degree categories: sparsely-rimed aggregates, moderately rimed, and rimed. Only shielded MASC measurements are shown.}
\label{fig:theta_chi}
\end{figure}

Distributions of unshielded MASC-measured orientation angles tend to favor high angles in high winds ($U_{sfc}>5\,\unit{m\,s^{-1}}$, Fig. \ref{fig:theta}(a)), where the mode is $57\unit{^{\circ}}$, but this shifts to $28\unit{^{\circ}}$ for the lightest winds ($U_{sfc}\leq1.5\,\unit{m\,s^{-1}}$, Fig. \ref{fig:theta}(c)). Shielded measurements tend towards even lower angles in the lightest winds, with a mode of $12\unit{^{\circ}}$ for $U_{sfc}\leq1.5\,\unit{m\,s^{-1}}$ (Fig. \ref{fig:theta}(f)). These results suggest that these solid hydrometeors tend to fall with their maximum dimensions nearly aligned with the horizontal plane when left undisturbed by surface winds. When separated by riming class for the lightest winds ($U_{sfc}\leq1.5\,\unit{m\,s^{-1}}$), shielded MASC orientation angles tend to be larger (i.e., more vertical) for sparsely-rimed aggregates (Fig. \ref{fig:theta_chi}). 

\begin{figure}[t]
\centering
\includegraphics[width=1\textwidth]{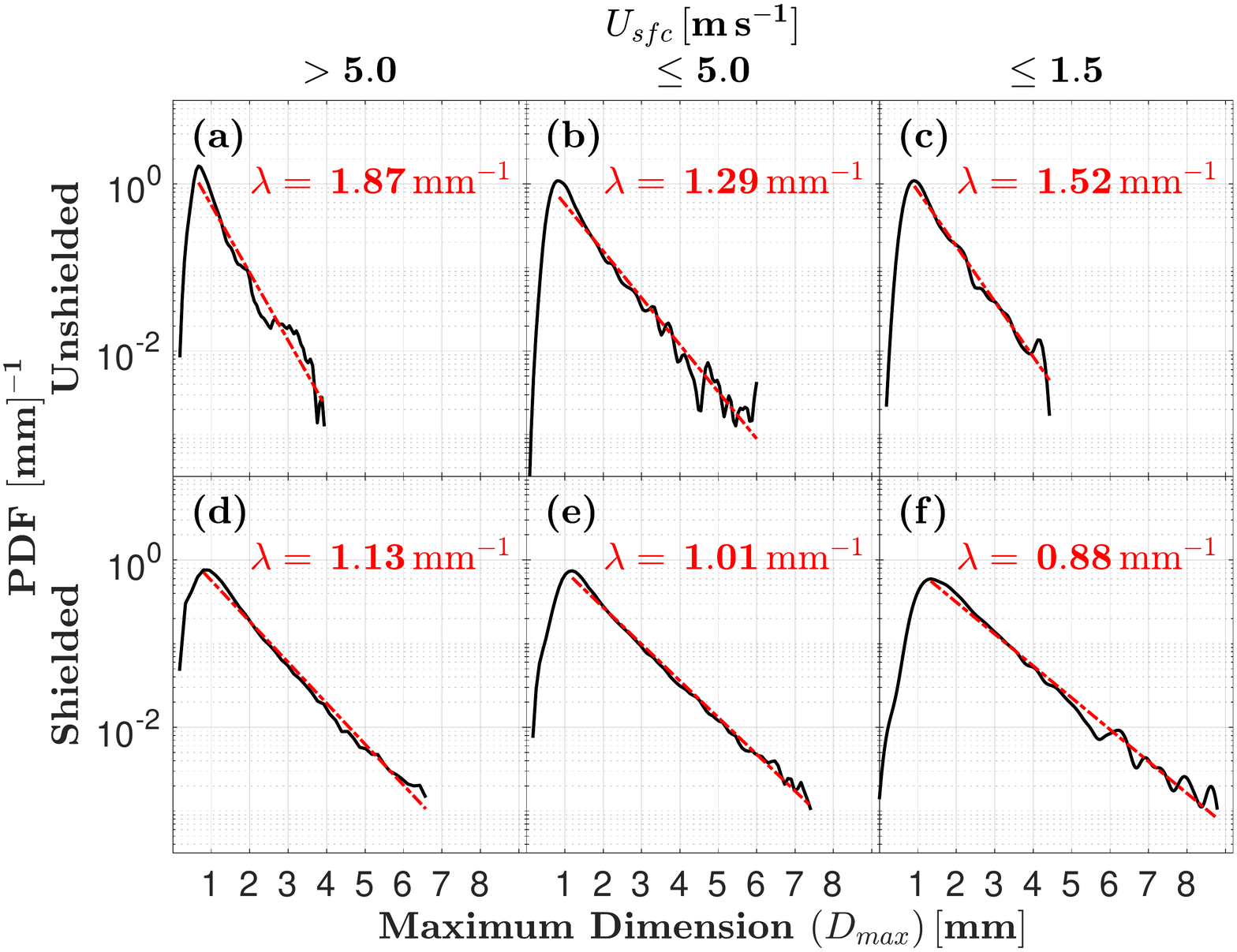}
\caption{As in Fig. \ref{fig:theta} but for maximum dimension $D_{max}$ and slope parameter $\lambda$. The slope parameter is calculated as the linear least-squares fit from the peak through the tail of the distribution.}
\label{fig:dmax}
\end{figure}

\begin{figure}[t]
\centering
\includegraphics[width=1\textwidth]{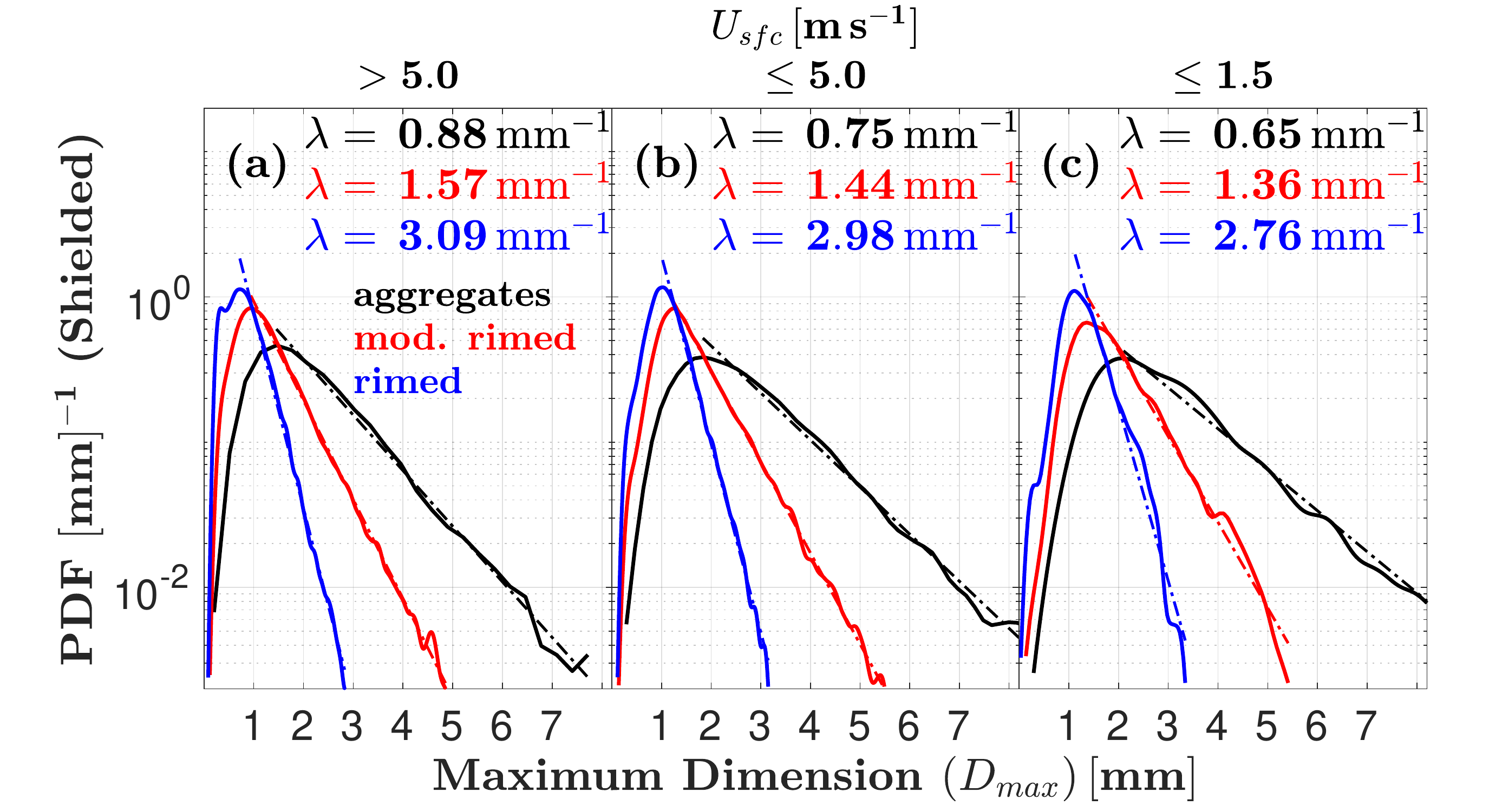}
\caption{As in Fig. \ref{fig:dmax} but with hydrometeors divided into three riming degree categories: sparsely-rimed aggregates, moderately rimed, and rimed. Only shielded MASC measurements are shown.}
\label{fig:dmax_chi}
\end{figure}

To examine surface wind influence on hydrometeor sizes observed by the MASC, distributions of $D_{max}$ and corresponding $\lambda$ values are shown in Fig. \ref{fig:dmax}. The slope parameter $\lambda$ is smallest when the MASC is shielded and surface winds are very light ($U_{sfc}\leq1.5\,\unit{m\,s^{-1}}$, Fig. \ref{fig:dmax}(f)), and largest for unshielded observations in high winds ($U_{sfc}>5\,\unit{m\,s^{-1}}$, Fig. \ref{fig:dmax}(a)). This suggests that the largest hydrometeors are less likely to be captured by the MASC in strong winds, and even less likely without shielding. When these wind-shielded distributions are separated into riming degree classes (Fig. \ref{fig:dmax_chi}), aggregates exhibit a 26\% percent decrease in $\lambda$, from 0.88 to 0.65 $\unit{mm^{-1}}$, when comparing high winds ($U_{sfc}>5\,\unit{m\,s^{-1}}$) to low winds ($U_{sfc}\leq1.5\,\unit{m\,s^{-1}}$). For a size distribution with the form $n_{D_{max}}=n_{D_0}\exp({-\lambda D_{max}})$, where $n_{D_{max}}\Delta D_{max}$ is the concentration of particles with sizes between $D_{max}$ and $D_{max}+\Delta D_{max}$, this decrease in $\lambda$ corresponds to a number concentration that is 5 times higher for aggregates with $D_{max}=7\,\unit{mm}\pm \Delta D_{max}/2$. In contrast, moderately and heavily rimed hydrometeors only exhibit a $\lambda$ decrease of 13\% and 11\%, respectively, when comparing high- and low-wind measurements.

The observation that measured concentrations of larger aggregates are relatively sensitive to surface winds compared to more heavily rimed particle types suggests that the frequency distribution of riming classes observed by the MASC might also reflect this sensitivity. Indeed, Table \ref{table:counts} shows that the percentage of wind-shielded aggregates reaches a maximum (25\%) when wind speeds are lowest ($U_{sfc}\leq0.5\,\unit{m\,s^{-1}}$). The opposite is true for shielded rimed hydrometeors (i.e., graupel), implying that high-density rimed particles are more likely to be observed by the MASC than large, weakly rimed aggregates in the presence of strong winds ($U_{sfc}>5\,\unit{m\,s^{-1}}$).


\begin{figure}[t]
\centering
\includegraphics[width=1\textwidth]{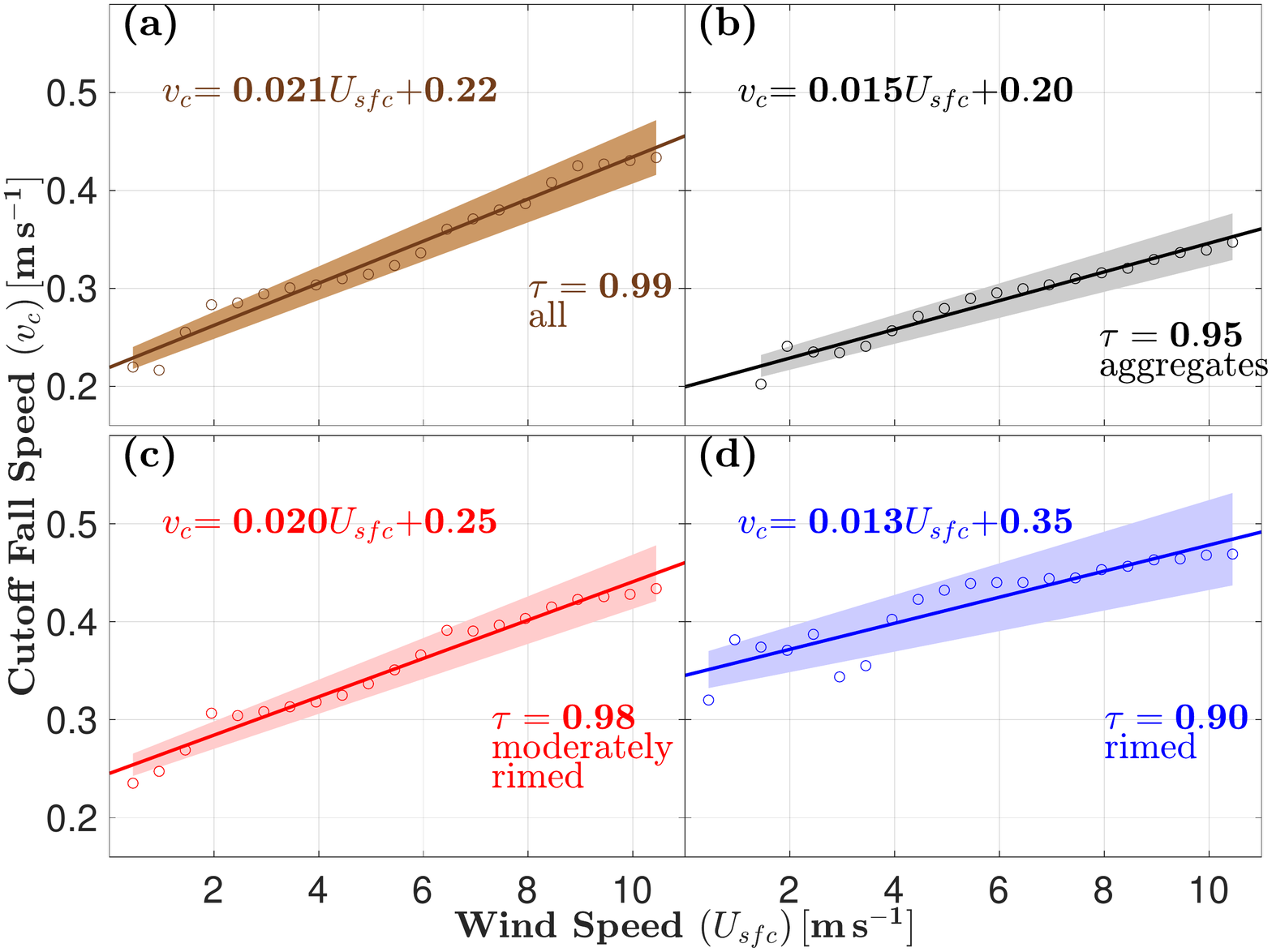}
\caption{Cutoff MASC fall speed $v_c$, defined in Sect. \ref{obs_fall_speed}, as a function of surface wind speed $U_{sfc}$ for (a) all hydrometeor types, (b) aggregates, (c) moderately rimed, and (d) rimed. The solid line in each subplot is a linear least squares best fit, while the shaded regions bound the 95\% confidence interval. Goodness-of-fit is measured by applying the Kendall rank correlation coefficient $\tau=2(P-Q)/n(n-1)$ \citep{kendall1938new}, where $P$ is the number of concordant pairs, $Q$ is the number of discordant pairs, and $n$ is the total number of pairs. A value of $\tau = 0$ indicates no relationship and 1 indicates a perfect relationship. The confidence interval represents the range of error for predicting a new value for $v_c$. Only shielded MASC measurements are shown.}
\label{vc_plot}
\end{figure}

\section{Discussion}
\label{discussion}

The cutoff fall speed $v_c$ defined in Sect. \ref{obs_fall_speed} is a potentially useful threshold for quality control of MASC fall speed measurements, and Fig. \ref{fig:v_vd_comp} suggests that $v_c=v_c(U_{sfc})$ for shielded MASC measurements. Least squares linear regression fits of $v_c$ to $U_{sfc}$ are plotted in Fig. \ref{vc_plot} in increments of 0.5 \unit{m\,s^{-1}}. Goodness-of-fit is 0.95 or greater for all but the most heavily rimed particles, where a value of 0 indicates no relationship, and 1 indicates a perfect relationship. Data points tend to fall outside the 95\% confidence interval for the most restricted wind speeds ($U_{sfc}<2\,\unit{m\,s^{-1}}$, or $<4\,\unit{m\,s^{-1}}$ for graupel), corresponding to the lowest number of observations. These fits can be used as a guide for quality control of shielded MASC measurements, where particles with fall speeds below $v_c$ are either omitted or corrected through extrapolation.

Larger aggregates with negligible riming tend to be more susceptible to disturbance by surface winds and associated turbulence, with more vertical orientations and lower frequency of occurrence than other riming classes. This finding supports that of \citet{Theriault_etal_2012}, who found that faster-falling hydrometeors are collected more efficiently by a Geonor gauge inside a single Alter shield. Therefore particle type needs to be considered when accounting for the effect of wind speed on snow measurements. However, the collection efficiencies for all riming classes sampled in the present study are found to be highly sensitive to winds in the absence of a wind shield. This sensitivity is reduced but still apparent for all but perhaps the very lightest winds $U_{sfc}\leq0.5\,\unit{m\,s^{-1}}$, even when located inside of a double wind fence. This is likely the result of upstream turbulence propagating into the collection area as a result of wind interacting with shield deflector fins, as suggested in \citet{colli2016collection1,colli2016collection2}. 

Prior MASC observations \citep{Garrett2014} and work by \citet{nielsen2007mean} have shown that fall speed distributions are broadened in highly turbulent flows, where fall speeds are either enhanced are reduced by turbulent eddies. Considering also that the MASC observes one hydrometeor at a time, while the KAZR fall speed is the mean value from a volume of scattering hydrometeors, it is certainly possible that at least some of the measurements comprising the low-fall-speed mode of the MASC fall speed distributions are a natural result of turbulence and not caused by the interaction of surface winds with the MASC or MASC-shield configuration. However, without more direct fall speed measurements to compare with, the highest confidence in the MASC fall speed measurements is achieved by omitting measured fall speeds that fall below $v_c$.


\conclusions  
\label{conclusions}
Accurate measurement of solid hydrometeor fall speed, orientation, and size distribution is critical for constraining numerical model parameterizations and remote sensing retrievals. Surface winds are known to have a strong influence on the collection of solid hydrometeors that is dependent on the specific gauge-shield configuration. Simulations of wind interactions with an unshielded MASC showed an average reduction in mean particle fall speed of 74\% for winds increasing to $10\,\unit{m\,s^{-1}}$, while TKE had only a weak, inverse effect on the reduction. In comparison with coincident KAZR observations of mean Doppler velocity, MASC measurements of fall speed were in closest agreement only when both the MASC was shielded with a double wind fence and winds were light ($U_{sfc}\leq5\,\unit{m\,s^{-1}}$). For the lightest wind speeds ($U_{sfc}\leq1.5\,\unit{m\,s^{-1}}$), shielded measurements of orientation angles decreased to a mode of $12\,\unit{^{\circ}}$, and concentrations of sparsely-rimed aggregates with $D_{max}\simeq7\,\unit{mm}$ increased by a factor of five. However, we showed that even in these wind-restricted and shielded cases, a fraction of MASC-measured fall speeds -- those below a wind-speed-dependent cutoff fall speed that is most often $v_c\lesssim0.5$ -- still do not match KAZR measurements. We showed that this cutoff fall speed is a function of wind speed for shielded observations and provided linear regression fits that can be used for additional quality control of MASC measurements. 

Future work could include a double wind fence in the CFD simulation to see more precisely how the wind field evolves as it encounters the individual deflector fins in each portion of the fence, where these fins are allowed to move with the wind. \citet{Theriault_etal_2012} simulated the wind field for a Geonor gauge with a single Alter shield by accounting for the movement of deflector fins on the upstream side of the gauge, where fins were assigned angles with respect to the vertical that increased as a function of wind speed. Such careful simulation might improve the fidelity of wind-shield-gauge influence on snow measurements.

The intent of this work is to provide guidance for under what measurement conditions the MASC can be used to obtain accurate information about hydrometeor microphysical properties and fall speeds. However, those conditions are limited to measurements within still air. The distributions of frozen hydrometeor size, type, orientation, and fall speed in natural, turbulent air remain to be determined. 





\codedataavailability{The code and data supporting this project are available at https://doi.org/10.7278/S50DQTX9K7QY. This repository includes code sufficient to replicate the observations analysis results. Raw and processed MASC data are available from the ARM data archive at https://adc.arm.gov/discovery/\#/, and raw MASC data can be processed with the \textit{mascpy} code located at https://doi.org/10.7278/S50DVA5JK2PD. OpenFoam v4.1 software can be downloaded at https://openfoam.org/version/4-1/.} 












\authorcontribution{All authors contributed to the formulation of the project. KF and TG developed the methodology and software code for observations analysis. CH developed the methodology and software implementation for the simulations, with AT advising. KF and CH wrote the article with contributions from TG and AT.} 

\competinginterests{TG is co-owner and scientific advisor of Particle Flux Analytics, Inc., the company manufacturing the Multi-Angle Snowflake Camera. Otherwise, the authors declare that they have no conflict of interest.} 


\begin{acknowledgements}
This work was supported by the Department of Energy Atmospheric System Research program Grant DE-SC0016282 and the National Science Foundation (NSF) Physical and Dynamic Meteorology program award number 1841870. We thank Krista Gaustad and Martin Stuefer for sharing precise dates, locations, and orientations of the instrument and wind shield.
\end{acknowledgements}







\bibliographystyle{copernicus}
\bibliography{references.bib}

\end{document}